\documentclass[traditabstract,rnote]{aa} 

\usepackage{array}
\usepackage{graphicx}
\usepackage{natbib}
\usepackage{lscape}
\usepackage{rotating}
\newcommand{\xmm}{{\sc XMM}\emph{-Newton}}
\newcommand{\ch}{\emph{Chandra}}

\begin{document}

\title{The modulated X-ray emission of the magnetic O8.5V-star Tr16-22\thanks{Based on observations collected with the ESA science mission {\it XMM-Newton}, {\it Chandra}, and ESO-FORS2 instrument.}}

\author{Ya\"el Naz\'e\inst{1}\fnmsep\thanks{FNRS Research Associate}, Gregg Wade\inst{2}, V\'eronique Petit\inst{3,4}}

\institute{Groupe d'Astrophysique des Hautes Energies, Institut d'Astrophysique et de G\'eophysique, Universit\'e de Li\`ege, 17, All\'ee du 6 Ao\^ut, B5c, B-4000 Sart Tilman, Belgium\\
\email{naze@astro.ulg.ac.be}
\and
Department of Physics, Royal Military College of Canada, PO Box 17000, Station Forces, Kingston, ON K7K 4B4, Canada
\and
Department of Physics \& Astronomy, University of Delaware, Bartol Research Institute, Newark, DE 19716, USA
\and
Dept. of Physics \& Space Sciences, Florida Institute of Technology, Melbourne, FL, 32901}

\authorrunning{Naz\'e et al.}
\titlerunning{X-rays from Tr16-22}

\abstract{Using an extensive X-ray dataset, we analyze the X-ray emission of the massive O-star Tr16-22, which was recently found to be magnetic. Its bright X-ray emission is found to be modulated with a $\sim$54d period. This timescale should represent the rotational timescale of the star, as for other magnetic massive stars. In parallel, new spectropolarimetric data confirm the published magnetic detection.}


   \maketitle

\section{Introduction}
After years of speculation, magnetic fields associated with O-stars were finally found in the last decade. While the usual Zeeman splitting cannot be easily detected in early-type stars because of their broad lines, the circular polarization associated with this effect is now detectable thanks to the advent of sensitive spectropolarimetric instruments. In total, several tens of OB stars are now known to be magnetic \citep[e.g.][]{pet13}. They share similar properties; and the picture that emerges is that of rare, strong, stable, and topologically simple magnetic fields. 

Such strong dipolar magnetic fields are able to guide the stellar winds from the two opposite hemispheres towards the magnetic equator where the flows collide \citep{bab97}. The gas is shock-heated to high temperatures, leading to X-ray emission in addition to the intrinsic high-energy emission of massive stars, and then cools to form a dense magnetosphere. In slowly-rotating O-type stars, the trapped material then falls back onto the star, or is ejected, forming a ``dynamical magnetosphere'' \citep[and references therein]{pet13,udd14}.

Tr16-22 displays a bright, variable, and hard X-ray emission atypical of single, ``normal'' massive stars \citep{eva04,ant08,naz11}. In view of its relatively late spectral type (O8.5V) and its high X-ray overluminosity (about 1\,dex), it is relatively unlikely, on both observational and theoretical grounds, that the high-energy emission of Tr16-22 arose in colliding-winds within a binary. Tr16-22 was thus considered as a candidate for magnetically confined winds. A spectropolarimetric investigation of such peculiar X-ray sources belonging to the Carina Nebula was performed by \citet{naz12carina}, and Tr16-22 was indeed found to be strongly magnetic. However, the scarce number of optical observations is currently insufficient to constrain further the properties of Tr16-22. In particular, one important characteristic, its rotational period, remains unknown though it is a crucial parameter when studying magnetic phenomena, notably magnetic braking \citep{udd09}. 

In this paper, we analyze an extensive series of X-ray observations of Tr16-22 to derive this parameter. Section 2 presents the observations used in the study, while Section 3 provides the results and Section 4 reports our conclusions.

\begin{table*}
 \centering
  \caption{List of the X-ray observations, ordered by observatory (second column) and exposure identifier (ObsID, quoted below in the third column along with the exposure time). Note that, for \xmm, there are several X-ray instruments which are switched on/off at different times: the quoted exposure times correspond to the shortest amongst them, usually that associated with the pn camera.}
  \label{list}
  \begin{tabular}{llccc}
  \hline
ID & Obs. & ObsID (exp. time) & Start Date & JD \\
  \hline
1 & XMM     & 0112560101 (23ks) & 2001-06-25@06:50:19 & 2452085.785\\
2 & XMM     & 0112560201 (24ks) & 2001-06-28@07:21:45 & 2452088.807\\
3 & XMM     & 0112560301 (29ks) & 2001-06-30@04:37:54 & 2452090.693\\
4 & XMM     & 0112580601 (28ks) & 2000-07-26@04:58:07 & 2451751.707\\
5 & XMM     & 0112580701 (8ks)  & 2000-07-27@23:48:14 & 2451753.492\\
6 & XMM     & 0145740101 (7ks)  & 2003-01-25@12:40:12 & 2452665.028\\
7 & XMM     & 0145740201 (7ks)  & 2003-01-27@00:45:34 & 2452666.532\\
8 & XMM     & 0145740301 (7ks)  & 2003-01-27@20:18:56 & 2452667.346\\
9 & XMM     & 0145740401 (8ks)  & 2003-01-29@01:22:23 & 2452668.557\\
10& XMM     & 0145740501 (7ks)  & 2003-01-29@23:36:58 & 2452669.484\\
11& XMM     & 0145780101 (8ks)  & 2003-07-22@01:33:19 & 2452842.565\\
12& XMM     & 0160160101 (15ks) & 2003-06-08@13:30:05 & 2452799.063\\
13& XMM     & 0160160901 (31ks) & 2003-06-13@23:33:50 & 2452804.482\\
14& XMM     & 0160560101 (12ks) & 2003-08-02@20:42:53 & 2452854.363\\
15& XMM     & 0160560201 (12ks) & 2003-08-09@01:25:59 & 2452860.560\\
16& XMM     & 0160560301 (19ks) & 2003-08-18@15:05:13 & 2452870.129\\
17& XMM     & 0311990101 (26ks) & 2006-01-31@17:46:03 & 2453767.240\\
18& XMM     & 0560580101 (14ks) & 2009-01-05@10:04:44 & 2454836.920\\
19& XMM     & 0560580201 (11ks) & 2009-01-09@14:09:49 & 2454841.090\\
20& XMM     & 0560580301 (26ks) & 2009-01-15@11:04:36 & 2454846.962\\
21& XMM     & 0560580401 (23ks) & 2009-02-02@04:28:00 & 2454864.686\\
22& XMM     & 0650840101 (27ks) & 2010-12-05@23:50:41 & 2455536.494\\
23& Chandra & 50 (12ks)$^a$     & 1999-09-06@19:48:12 & 2451428.325\\
24& Chandra & 632 (90ks)        & 2000-11-19@02:46:40 & 2451867.616\\
25& Chandra & 1249 (10ks)$^a$   & 1999-09-06@23:45:34 & 2451428.490\\
26& Chandra & 6402 (87ks)       & 2006-08-30@18:38:27 & 2453978.277\\
27& Chandra & 11993 (44ks)      & 2010-11-14@13:32:37 & 2455515.064\\
28& Chandra & 11994 (39ks)      & 2010-11-21@07:25:34 & 2455521.809\\
\hline
\end{tabular}
\\
Tr16-22 is out of the field-of-view for the other Carina observations.\\
Note: $^a$ No correct calibration could be calculated for these two observations
\end{table*}

\section{Observations}

\subsection{X-ray data}
Tr16-22 is the most frequently observed magnetic massive star, as it is located close to $\eta$\,Carinae, which was the subject of several intense monitoring by modern X-ray facilities. Twenty-eight X-ray observations are available, 22 by \xmm\ and 6 by \ch\ (Table \ref{list}). 

\xmm\ data were reduced with SAS v13.0.0 using calibration files available in June 2013 and following the recommendations of the \xmm\ team\footnote{SAS threads, see \\ http://xmm.esac.esa.int/sas/current/documentation/threads/ }. Only best-quality data ($PATTERN$ of 0--12 for MOS and 0--4 for pn) were kept and background flares were discarded. A source detection was performed on each EPIC dataset using the task {\it edetect\_chain} on the 0.4--10.0\,keV energy band and for a likelihood of 10, to find the best-fit position of the target in each dataset. Note that the source is not bright enough to present pile-up. EPIC spectra were extracted using the task {\it especget} for circular regions centered on these best-fit positions and, for the backgrounds, at positions as close as possible to the target considering crowding and CCD edges. The spectra were finally grouped, using {\it specgroup}, to obtain an oversampling factor of five and to ensure that a minimum signal-to-noise ratio of three (i.e. a minimum of 10 counts) was reached in each spectral bin of the background-corrected spectra.

The \textit{Chandra} ACIS observations were reprocessed following the standard reduction procedure with CIAO v4.5\footnote{CIAO threads see http://cxc.harvard.edu/ciao/threads/}. The procedure is similar to the one described for EPIC observations: source search (using \textit{celldetect}),  extraction of spectra (using \textit{specextract}), grouping of spectra. Note that two \ch\ observations (ObsID 50 and 1249) could not be correctly calibrated, and were thus discarded from our analysis. 

The X-ray spectra were then fitted within Xspec v12.7.0 using absorbed optically-thin thermal plasma models, i.e. $wabs \times phabs \times \sum apec$, with solar abundances \citep{and89}. The first absorption component is the interstellar column, fixed to $4.4\times10^{21}$\,cm$^{-2}$ (a value calculated using the color excess of the star and the conversion formula $5.8\times10^{21}\times E(B-V)$\,cm$^{-2}$ from \citealt{boh78}), while the second absorption represents additional (local) absorption. At first, since one thermal component was not enough to achieve a good fitting, we have considered two thermal components, and fitted all data independently with this model. Since the additional absorption and temperatures were not significantly varying amongst datasets, they were fixed to $6.3\times10^{21}$\,cm$^{-2}$, 0.28\,keV and 1.78\,keV, respectively, and this new model was used to re-fit all data. In addition, in a general survey of all magnetic stars (Naz\'e et al., submitted), we used a given set of four emission components with temperatures fixed to 0.2, 0.6, 1.0, and 4.0\,keV. We also applied this model to Tr16-22, and found that the additional absorption was not significantly varying and the emission measure of the second temperature was extremely small and always compatible with zero. They were thus fixed to $7\times10^{21}$\,cm$^{-2}$ and zero, respectively, and this modified model was then used for a last fitting run. It should be noted that both procedures yield the same results, within the errors. From the best-fit spectral parameters, we derived two indicators of the spectral shape: a hardness ratio, calculated as the ratio between the hard (2.--10.0\,keV) and soft (0.5--2.0\,keV) ISM absorption-corrected fluxes, and an average temperature, defined as $\overline{kT} = (\sum kT_i \times norm_i)/(\sum norm_i)$, where $norm_i$ are the normalization factors of the spectral fits. Note that the latter is more uncertain than the former, because of the usual absorption/temperature trade-off in the X-ray domain (for more details see Naz\'e et al., submitted). Tables \ref{2tfits} and \ref{4tfits} yield the results of our X-ray fittings. 

Previously, X-ray variations of this star have been reported by \citet{com11}. The latter authors, using seven of the numerous \xmm\ observations, reported variations in the absorption-corrected flux by a factor of 30, as well as changes in temperature and absorption. This may seem at odds with our results reported in next section, but there are several analysis differences explaining this. First, compared to our study, \citet{com11} did not try to avoid erratic results. For example, the reported parameters show the usual interplay between temperature and absorption (fits with lower absorption also have higher temperatures), which hides the true stellar variations. Moreover, their absorption-corrected fluxes were corrected by the {\it total} columns, instead of by only the interstellar ones as should be done for massive stars. In view of the (large) uncertainties on the additional absorption, such corrections often lead to values which thus are, to say the least, difficult to interpret. We therefore conclude that order-of-magnitude flux variations have not been observed in Tr16-22, nor large changes in temperature or absorption. 

\begin{figure}
\includegraphics[width=8.5cm]{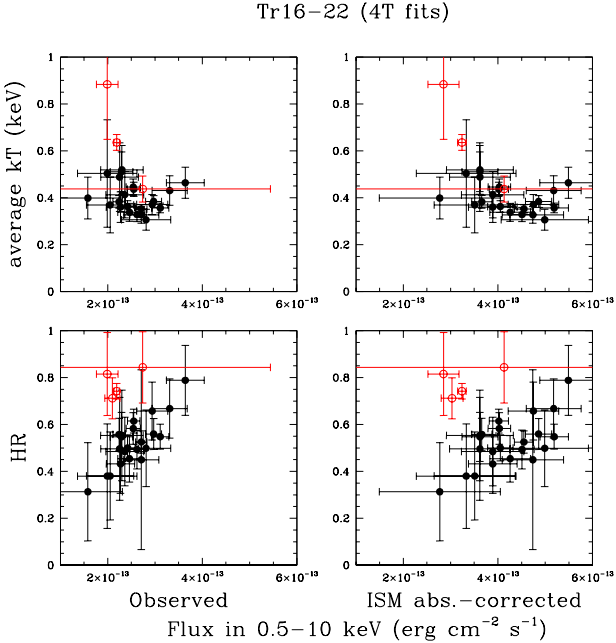}
\caption{Evolution of the average temperatures and hardness ratios ($HR=H/S=F^{ISMcor}(2.-10.0\,keV)/F^{ISMcor}(0.5-2.\,keV)$) as a function of observed and absorption-corrected fluxes. Red open symbols correspond to \ch\ data, while black filled symbols correspond to results from \xmm\ observations. The values shown here correspond to the results from 4T fits (results are similar for the 2T fits). }
\label{variab}
\end{figure}

\subsection{Spectropolarimetry}
Tr16-22 was again observed with FORS2 at the Very Large Telescope in 2013 on April 18 and July 29 (ESO ObsID=091.D-0090(B), PI Naz\'e, service mode - only two datasets of the planned monitoring could be obtained). The data were taken with the red CCD (a mosaic composed of two 2k$\times$4k MIT chips) without binning, a slit of 1'', and the 1200B grating ($R\sim 1400$). The observing sequence consisted of 8 subexposures of 240s duration with retarder waveplate positions of $+45^{\circ}$, $+45^{\circ}$, $-45^{\circ}$, $-45^{\circ}$, $+45^{\circ}$, $+45^{\circ}$, $-45^{\circ}$, $-45^{\circ}$. We reduced these spectropolarimetric data with IRAF as explained by \citet{naz12carina}: aperture extraction radius fixed to 20\,px, subtraction of nearby sky background, and wavelength calibration from 3675 to 5128\AA\ (with pixels of 0.25\AA) considering arc lamp data taken at only one retarder waveplate position (in our case, $+45^\circ$). This allowed us to construct the normalized Stokes $V/I$ profile, as well as a diagnostic ``null'' profile \citep{don97,bag09}. Finally, the associated longitudinal magnetic field was estimated by minimizing $\chi^2 = \sum_i \frac{(y_i - B_z\,x_i - a)^2}{\sigma^2_i}$ with $y_i$ either $V/I$ or the null profile at the wavelength $\lambda_i$ and $x_i = -g_\mathrm{eff}\ 4.67 \times 10^{-13} \ \lambda^2_i\ 1/I_i\ (\mathrm{d}I/\mathrm{d}\lambda)_i$ \citep{bag02}. This was done after discarding edges and deviant points, after rectifying the Stokes profiles, and after selecting spectral windows centered on lines \citep[see][for further discussion]{naz12carina}.

\begin{figure*}
\includegraphics[width=6cm]{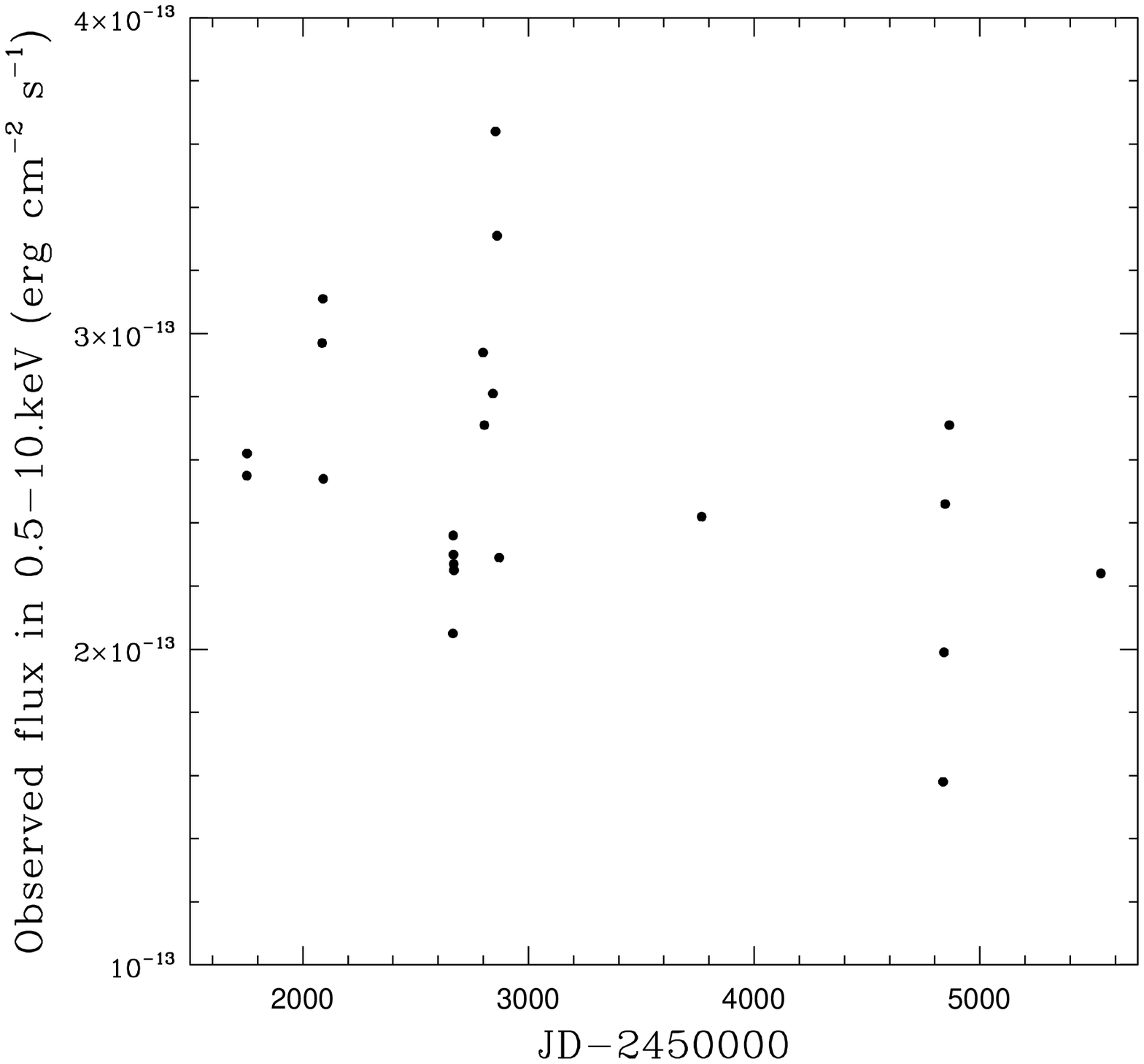}
\includegraphics[width=6cm]{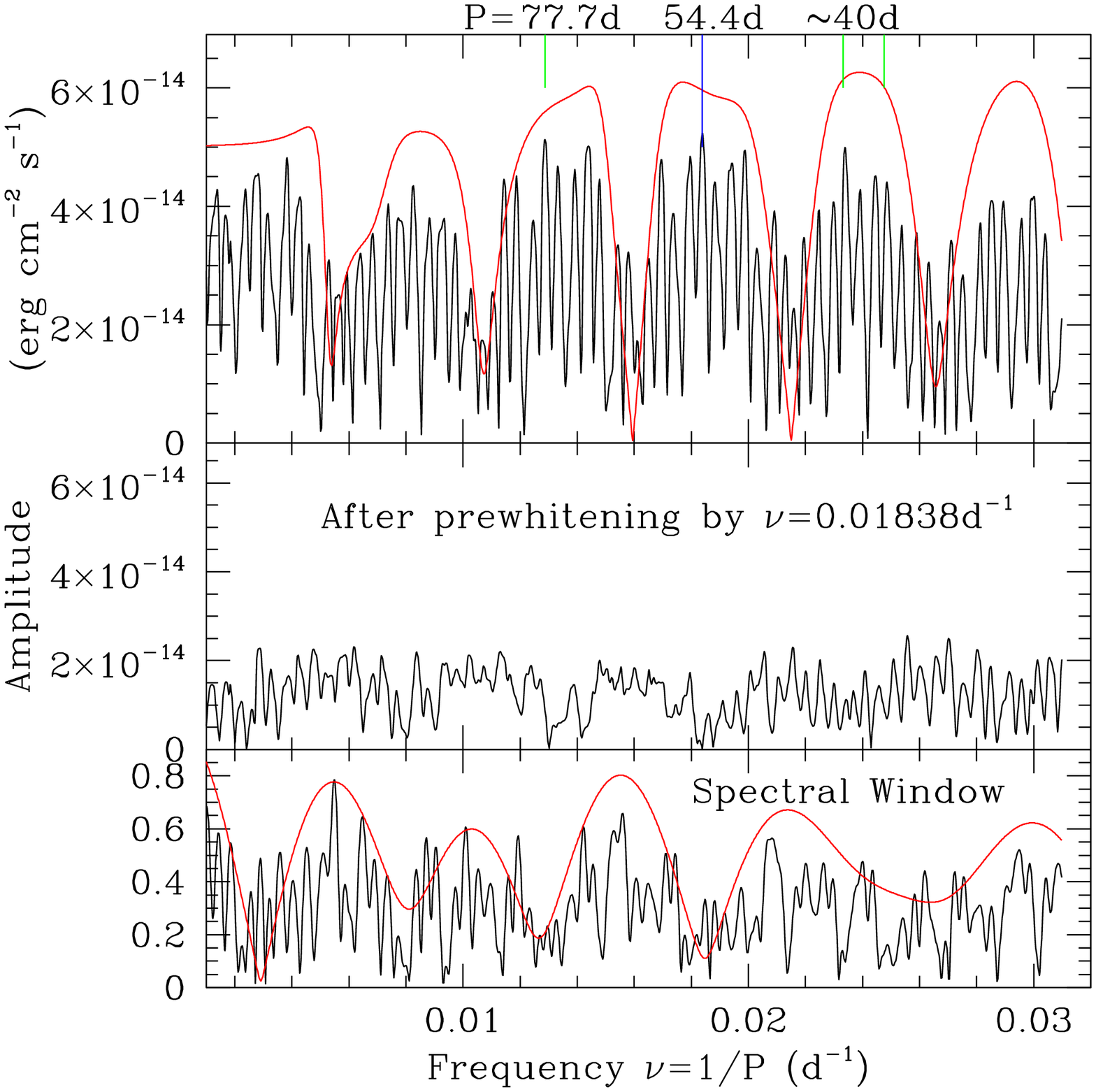}
\includegraphics[width=6cm]{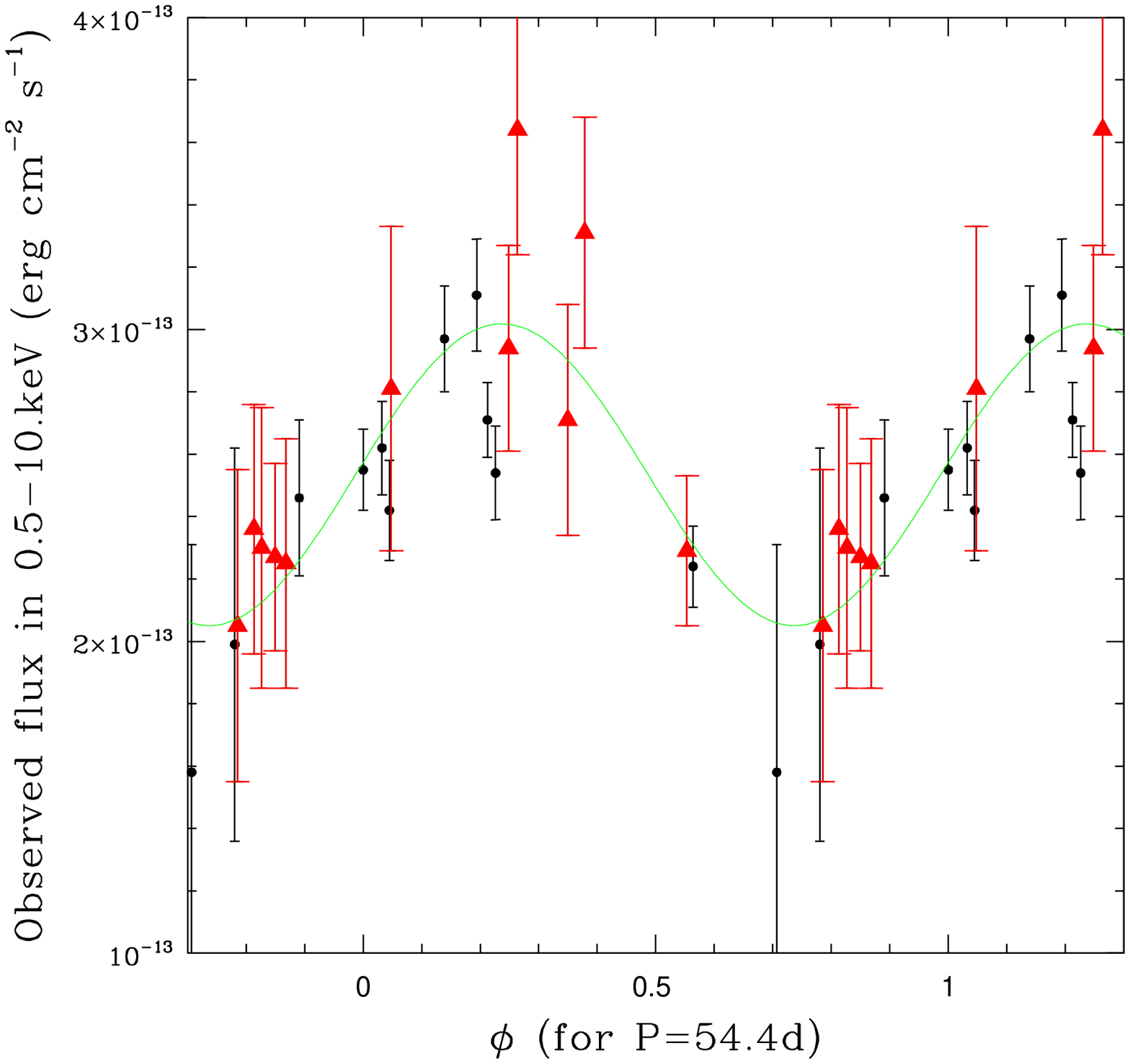}
\caption{{\it Left:} Evolution of the observed \xmm\ fluxes of Tr16-22 with time (for the 4T fits). {\it Middle:} Periodogram and spectral window for the whole (black curve) and 2003 (red curve) datasets of these fluxes. {\it Right:} Evolution of these fluxes when phasing with a frequency of 0.01838\,d$^{-1}$. Red triangles correspond to the 2003 data and the green line to the best-fit sinusoid.}
\label{tr1622}
\end{figure*}

\begin{table*}
  \caption{Results from the spectral fits, for the models $wabs \times phabs \times \sum_2 apec$ with absorptions fixed to $4.4\times10^{21}$\,cm$^{-2}$ and $6.3\times10^{21}$\,cm$^{-2}$, and temperatures fixed to 0.28\,keV and 1.78\,keV. The remaining free parameters, the strength of the two thermal components, are listed in columns 2 and 3, along with the goodness-of-fit and number of degrees of freedom (column 4). In columns 5 and 6, observed fluxes and ISM-absorption corrected luminosities are given in the 0.5--10.\,keV range and for a distance of 2290pc. The last two columns yield the average temperatures and hardness ratios (see text for definition).}
  \label{2tfits}
  \begin{tabular}{lccccccc}
  \hline
ID & $norm_1$ & $norm_2$ & $\chi^2$(dof) & $F_{\rm X}^{obs}$         & $L_{\rm X}^{ISMcor}$ & $kT_{avg}$ & HR \\
   & \multicolumn{2}{c}{(cm$^{-5}$)}&    & erg\,cm$^{-2}$\,s$^{-1}$ & erg\,s$^{-1}$     & keV       &    \\
  \hline
  \hline
 1 & 1.69$\pm$0.18e-03 & 3.50$\pm$0.19e-04 & 1.(76)    & 2.72$\pm$0.10e-13 & 2.92e+32 & 0.54$\pm$0.01 & 0.472$\pm$0.034  \\ 
 2 & 1.72$\pm$0.17e-03 & 3.57$\pm$0.19e-04 & 0.93(85)  & 2.77$\pm$0.07e-13 & 2.98e+32 & 0.54$\pm$0.01 & 0.472$\pm$0.029  \\ 
 3 & 1.34$\pm$0.15e-03 & 3.23$\pm$0.17e-04 & 1.18(79)  & 2.42$\pm$0.08e-13 & 2.53e+32 & 0.57$\pm$0.02 & 0.517$\pm$0.033  \\ 
 4 & 1.27$\pm$0.16e-03 & 3.18$\pm$0.16e-04 & 0.92(68)  & 2.36$\pm$0.09e-13 & 2.45e+32 & 0.58$\pm$0.02 & 0.533$\pm$0.039  \\ 
 5 & 1.63$\pm$0.25e-03 & 2.88$\pm$0.29e-04 & 1.29(33)  & 2.34$\pm$0.14e-13 & 2.60e+32 & 0.51$\pm$0.02 & 0.424$\pm$0.057  \\ 
 6 & 1.63$\pm$0.41e-03 & 2.53$\pm$0.46e-04 & 1.13(9)   & 2.14$\pm$0.22e-13 & 2.43e+32 & 0.48$\pm$0.04 & 0.392$\pm$0.062  \\ 
 7 & 1.51$\pm$0.41e-03 & 2.99$\pm$0.50e-04 & 1.61(12)  & 2.35$\pm$0.25e-13 & 2.55e+32 & 0.53$\pm$0.04 & 0.464$\pm$0.090  \\ 
 8 & 1.32$\pm$0.42e-03 & 3.06$\pm$0.48e-04 & 0.55(9)   & 2.31$\pm$0.23e-13 & 2.44e+32 & 0.56$\pm$0.04 & 0.510$\pm$0.109  \\ 
 9 & 1.60$\pm$0.35e-03 & 2.83$\pm$0.40e-04 & 1.15(16)  & 2.30$\pm$0.20e-13 & 2.55e+32 & 0.51$\pm$0.03 & 0.430$\pm$0.078  \\ 
10 & 1.45$\pm$0.41e-03 & 3.02$\pm$0.48e-04 & 1.19(12)  & 2.34$\pm$0.23e-13 & 2.52e+32 & 0.54$\pm$0.04 & 0.480$\pm$0.097  \\ 
11 & 1.67$\pm$5.15e-03 & 2.91$\pm$0.58e-04 & 0.95(7)   & 2.37$\pm$0.28e-13 & 2.64e+32 & 0.50$\pm$0.12 & 0.425$\pm$0.100  \\ 
12 & 1.25$\pm$0.34e-03 & 3.40$\pm$0.43e-04 & 0.55(15)  & 2.48$\pm$0.20e-13 & 6.25e+31 & 0.60$\pm$0.04 & 0.560$\pm$0.083  \\ 
13 & 1.86$\pm$0.30e-03 & 3.03$\pm$0.34e-04 & 1.47(19)  & 2.52$\pm$0.18e-13 & 2.84e+32 & 0.49$\pm$0.02 & 0.407$\pm$0.051  \\ 
14 & 1.36$\pm$0.44e-03 & 4.36$\pm$0.50e-04 & 0.66(14)  & 3.08$\pm$0.25e-13 & 3.06e+32 & 0.64$\pm$0.04 & 0.614$\pm$0.092  \\ 
15 & 1.45$\pm$0.41e-03 & 4.07$\pm$0.47e-04 & 1.10(16)  & 2.95$\pm$0.26e-13 & 3.00e+32 & 0.61$\pm$0.04 & 0.572$\pm$0.085  \\ 
16 & 1.34$\pm$0.27e-03 & 2.99$\pm$0.32e-04 & 0.56(19)  & 2.28$\pm$0.16e-13 & 2.42e+32 & 0.55$\pm$0.03 & 0.498$\pm$0.073  \\ 
17 & 1.47$\pm$0.16e-03 & 2.96$\pm$0.18e-04 & 1.05(50)  & 2.31$\pm$0.09e-13 & 2.50e+32 & 0.53$\pm$0.02 & 0.458$\pm$0.036  \\ 
18 & 1.51$\pm$0.39e-03 & 1.95$\pm$0.47e-04 & 1.58(6)   & 1.75$\pm$0.20e-13 & 2.07e+32 & 0.45$\pm$0.04 & 0.347$\pm$0.096  \\ 
19 & 1.50$\pm$0.45e-03 & 2.86$\pm$0.54e-04 & 1.14(8)   & 2.27$\pm$0.26e-13 & 2.49e+32 & 0.52$\pm$0.05 & 0.451$\pm$0.099  \\ 
20 & 1.67$\pm$0.28e-03 & 2.84$\pm$0.32e-04 & 0.71(22)  & 2.33$\pm$0.16e-13 & 2.60e+32 & 0.50$\pm$0.03 & 0.420$\pm$0.059  \\ 
21 & 1.57$\pm$0.14e-03 & 3.15$\pm$0.17e-04 & 0.88(74)  & 2.47$\pm$0.08e-13 & 2.67e+32 & 0.53$\pm$0.01 & 0.460$\pm$0.030  \\ 
22 & 1.27$\pm$0.14e-03 & 2.64$\pm$0.18e-04 & 1.16(51)  & 2.05$\pm$0.10e-13 & 2.20e+32 & 0.54$\pm$0.02 & 0.475$\pm$0.034  \\ 
24 & 6.35$\pm$1.47e-04 & 3.17$\pm$0.22e-04 & 1.08(39)  & 2.09$\pm$0.10e-13 & 1.95e+32 & 0.78$\pm$0.03 & 0.761$\pm$0.069  \\ 
26 & 8.74$\pm$0.64e-04 & 2.99$\pm$0.10e-04 & 0.98(130) & 2.09$\pm$0.05e-13 & 2.06e+32 & 0.66$\pm$0.01 & 0.640$\pm$0.022  \\ 
27 & 6.22$\pm$2.70e-04 & 2.83$\pm$0.32e-04 & 0.83(18)  & 1.89$\pm$0.11e-13 & 1.79e+32 & 0.75$\pm$0.05 & 0.732$\pm$0.140  \\ 
28 & 8.94$\pm$3.44e-04 & 3.41$\pm$0.41e-04 & 1.10(14)  & 2.34$\pm$0.22e-13 & 1.36e+32 & 0.69$\pm$0.05 & 0.671$\pm$0.124  \\ 
\hline
\end{tabular}
\end{table*}

\begin{table*}
  \caption{Same as Table \ref{2tfits} but for the models $wabs \times phabs \times \sum_4 apec$ with absorptions fixed to $4.4\times10^{21}$\,cm$^{-2}$ and $7.\times10^{21}$\,cm$^{-2}$, and temperatures fixed to 0.2, 0.6, 1.0, and 4.0\,keV, and a zero normalization factor $norm_2$. }
  \label{4tfits}
  \begin{tabular}{lcccccccc}
  \hline
ID & $norm_1$ & $norm_3$ & $norm_4$ & $\chi^2$(dof) & $F_{\rm X}^{obs}$         & $L_{\rm X}^{ISMcor}$ & $kT_{avg}$ & HR \\
   & \multicolumn{3}{c}{(cm$^{-5}$)}&               & erg\,cm$^{-2}$\,s$^{-1}$ & erg\,s$^{-1}$     & keV       &    \\
  \hline
  \hline
 1 &  3.62$\pm$0.95e-03 &  3.43$\pm$0.45e-04 & 1.26$\pm$0.22e-04 & 0.93(75)  & 2.97$\pm$0.17e-13 & 3.05e+32& 0.38$\pm$0.03 & 0.559$\pm$0.066  \\ 
 2 &  4.59$\pm$0.91e-03 &  3.07$\pm$0.44e-04 & 1.43$\pm$0.22e-04 & 1.00(84)  & 3.11$\pm$0.18e-13 & 3.26e+32& 0.36$\pm$0.02 & 0.548$\pm$0.053  \\ 
 3 &  2.19$\pm$0.82e-03 &  3.46$\pm$0.39e-04 & 9.75$\pm$1.92e-05 & 1.16(78)  & 2.54$\pm$0.15e-13 & 2.53e+32& 0.45$\pm$0.03 & 0.583$\pm$0.067  \\ 
 4 &  2.39$\pm$0.82e-03 &  3.12$\pm$0.37e-04 & 1.09$\pm$0.17e-04 & 1.05(67)  & 2.55$\pm$0.13e-13 & 2.52e+32& 0.44$\pm$0.03 & 0.614$\pm$0.051  \\ 
 5 &  4.61$\pm$1.31e-03 &  2.48$\pm$0.66e-04 & 1.16$\pm$0.33e-04 & 1.31(32)  & 2.62$\pm$0.15e-13 & 2.83e+32& 0.33$\pm$0.03 & 0.493$\pm$0.083  \\ 
 6 &  2.12$\pm$2.98e-03 &  3.83$\pm$1.63e-04 & 3.25$\pm$6.57e-05 & 1.12(8)   & 2.05$\pm$0.50e-13 & 2.20e+32& 0.37$\pm$0.12 & 0.380$\pm$0.187  \\ 
 7 &  2.17$\pm$2.42e-03 &  3.79$\pm$1.25e-04 & 6.69$\pm$5.60e-05 & 1.60(11)  & 2.36$\pm$0.40e-13 & 2.44e+32& 0.41$\pm$0.10 & 0.485$\pm$0.147  \\ 
 8 &  1.32$\pm$2.58e-03 &  3.85$\pm$1.37e-04 & 6.79$\pm$0.60e-05 & 0.47(8)   & 2.30$\pm$0.45e-13 & 2.27e+32& 0.52$\pm$0.10 & 0.554$\pm$0.193  \\ 
 9 &  2.80$\pm$2.00e-03 &  3.53$\pm$1.01e-04 & 6.11$\pm$4.58e-05 & 1.29(15)  & 2.27$\pm$0.30e-13 & 2.44e+32& 0.36$\pm$0.07 & 0.432$\pm$0.126  \\ 
10 &  1.38$\pm$2.49e-03 &  4.07$\pm$1.27e-04 & 5.40$\pm$5.61e-05 & 1.25(11)  & 2.25$\pm$0.40e-13 & 2.27e+32& 0.49$\pm$0.15 & 0.496$\pm$0.219  \\ 
11 &  6.27$\pm$2.97e-03 &  1.65$\pm$1.38e-04 & 1.49$\pm$0.64e-04 & 1.03(6)   & 2.81$\pm$0.52e-13 & 3.13e+32& 0.31$\pm$0.04 & 0.498$\pm$0.163  \\ 
12 &  4.29$\pm$1.86e-03 &  2.14$\pm$0.90e-04 & 1.65$\pm$0.43e-04 & 0.64(14)  & 2.94$\pm$0.33e-13 & 8.35e+31& 0.37$\pm$0.04 & 0.657$\pm$0.124  \\ 
13 &  4.57$\pm$1.65e-03 &  3.14$\pm$0.79e-04 & 1.01$\pm$0.39e-04 & 1.71(19)  & 2.71$\pm$0.37e-13 & 2.97e+32& 0.33$\pm$0.04 & 0.450$\pm$0.383  \\ 
14 &  3.39$\pm$2.33e-03 &  3.17$\pm$1.16e-04 & 2.05$\pm$0.55e-04 & 0.50(13)  & 3.64$\pm$0.40e-13 & 3.45e+32& 0.46$\pm$0.07 & 0.788$\pm$0.149  \\ 
15 &  3.35$\pm$2.30e-03 &  3.48$\pm$1.12e-04 & 1.61$\pm$0.51e-04 & 1.21(15)  & 3.31$\pm$0.37e-13 & 3.25e+32& 0.43$\pm$0.06 & 0.668$\pm$0.126  \\ 
16 &  1.35$\pm$1.42e-03 &  3.86$\pm$0.70e-04 & 6.63$\pm$3.00e-05 & 0.50(18)  & 2.29$\pm$0.24e-13 & 2.27e+32& 0.51$\pm$0.08 & 0.547$\pm$0.105  \\ 
17 &  3.23$\pm$0.82e-03 &  2.98$\pm$0.40e-04 & 9.22$\pm$1.95e-05 & 1.34(49)  & 2.42$\pm$0.16e-13 & 2.54e+32& 0.36$\pm$0.02 & 0.500$\pm$0.056  \\ 
18 &  1.15$\pm$1.91e-03 &  3.80$\pm$0.73e-04 & 0.00$\pm$2.18e-05 & 1.27(5)   & 1.58$\pm$0.73e-13 & 1.74e+32& 0.40$\pm$0.09 & 0.313$\pm$0.209  \\ 
19 &  8.97$\pm$28.4e-04 &  4.64$\pm$1.51e-04 & 1.21$\pm$6.48e-05 & 1.26(7)   & 1.99$\pm$0.63e-13 & 2.09e+32& 0.50$\pm$0.23 & 0.380$\pm$0.223  \\ 
20 &  3.76$\pm$1.52e-03 &  3.11$\pm$0.72e-04 & 8.54$\pm$3.51e-05 & 0.74(21)  & 2.46$\pm$0.25e-13 & 2.67e+32& 0.34$\pm$0.04 & 0.454$\pm$0.099  \\ 
21 &  4.02$\pm$0.75e-03 &  2.83$\pm$0.36e-04 & 1.18$\pm$0.17e-04 & 1.04(73)  & 2.71$\pm$0.12e-13 & 2.86e+32& 0.35$\pm$0.02 & 0.525$\pm$0.048  \\ 
22 &  2.74$\pm$0.73e-03 &  2.60$\pm$0.39e-04 & 9.40$\pm$1.80e-05 & 1.1(50)   & 2.24$\pm$0.13e-13 & 2.30e+32& 0.38$\pm$0.03 & 0.557$\pm$0.069  \\ 
24 &  0.00$\pm$4.36e-04 &  3.66$\pm$0.33e-04 & 6.88$\pm$1.47e-05 & 1.03(38)  & 2.10$\pm$0.16e-13 & 1.90e+32& 1.47$\pm$0.30 & 0.712$\pm$0.088  \\ 
26 &  9.83$\pm$3.24e-04 &  2.93$\pm$0.19e-04 & 9.56$\pm$0.89e-05 & 1.11(130) & 2.19$\pm$0.06e-13 & 2.03e+32& 0.64$\pm$0.03 & 0.742$\pm$0.033  \\ 
27 &  4.49$\pm$14.3e-04 &  2.74$\pm$0.68e-04 & 8.83$\pm$2.47e-05 & 0.86(17)  & 1.99$\pm$0.23e-13 & 1.79e+32& 0.88$\pm$0.23 & 0.815$\pm$0.177  \\ 
28 &  3.02$\pm$1.86e-03 &  1.73$\pm$0.93e-04 & 1.74$\pm$0.38e-04 & 1.03(13)  & 2.74$\pm$2.70e-13 & 1.41e+32& 0.44$\pm$0.06 & 0.844$\pm$0.152  \\ 
\hline
\end{tabular}
\end{table*}

\section{Results}
It is clear that Tr16-22 has a variable X-ray emission. The fluxes and the hardness ratios double while the average temperature varies by 40\% (Fig. \ref{variab} and Tables \ref{2tfits} \& \ref{4tfits}). The trend of a higher flux with a higher hardness ratio is best seen on \xmm\ data, because of their higher quality, their greater number, and the fact that they cover a large range of flux and hardness values. On the contrary, \ch\ data, fewer in number and of lower quality, are gathered around a specific value of flux/hardness and do not span the whole flux/hardness range, thereby rendering the trend more difficult to see. We also note that the \ch\ data yield higher temperatures and hardness ratios, as expected from the known calibration differences \citep{sch13}. 

The large number of X-ray data enabled us to examine these variations in details, in particular as a function of time (see left panel of Fig. \ref{tr1622}). To keep the dataset homogeneous, we concentrated on the \xmm\ data, because of their greater number and of the calibration difference between the two observatories. In this context, it should be noted that observations of Tr16-22 were taken between 2000 and 2010: changes due to aging detectors could be expected over such a long interval of time but only a few percent decrease in broad-band fluxes may be expected for aging effects not already corrected by the current \xmm\ calibration \citep{naz12zeta,naz13zeta}, i.e. much less than the observed amplitude of the changes. In addition, since half of the exposures were taken in 2003, all conclusions can be easily checked against such effects (which would have no influence over the course of a year). 

Keeping this in mind, it seems reasonable to search for periodicities, since the variations of the magnetic O-stars HD\,191612 and $\theta^1$\,Ori\,C are periodic in the X-ray domain as well as in the optical \citep{gag05,naz10}. Such variations are interpreted as periodic occultations of the X-ray emitting confined winds by the stellar body in a magnetic oblique rotator configuration. We thus analyzed the observed \xmm\ fluxes (from the 4T fits) using a Fourier algorithm adapted to sparse datasets \citep{hmm,gos01}. Fig. \ref{tr1622} shows the evolution of the fluxes with time and the periodogram. The Fourier periodogram is not similar to the spectral window: some additional signal is present, but also its aliases due to the small number of observations and coarse sampling of the variations. The highest peak (hence statistically the best one) of the periodogram calculated for the full dataset occurs at a frequency of 0.01838\,d$^{-1}$ (or a period of 54.42$\pm$0.06\,d, see central panel of Fig. \ref{tr1622}). Similar results are found when only considering the 2003 data or when analyzing the results from 2T fits. If the best-fit sinusoid with that 54.4d period is taken out of the data, only low-level noise remains (see central panel of Fig. \ref{tr1622}). In addition, this period provides a very coherent phased diagram (see right panel of Fig. \ref{tr1622}). However, we caution that many aliases are present in the periodogram, as the sampling of the period by the X-ray data is  far from perfect. In this context, alternative periods (e.g. at 40.39d, 42.88d, 77.70d, see central panel of Fig. \ref{tr1622}) provide fits of similar, though slightly less good, quality to the data as well as less convincing phased variations.  Any of these relatively long periods, interpreted as the stellar rotation period, is in qualitative agreement with the observed narrowness of the photospheric lines \citep{naz12carina}. 

\begin{figure}
\includegraphics[width=8.5cm]{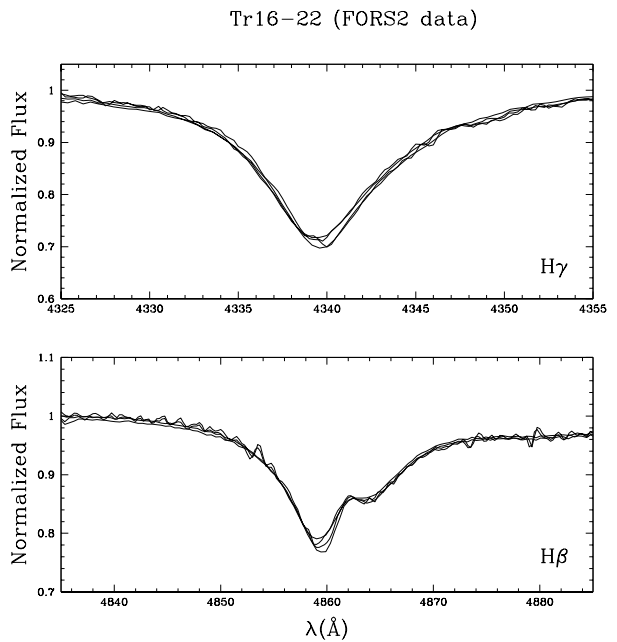}
\caption{H$\beta$ and H$\gamma$ lines in the four FORS2 observations.}
\label{balmer}
\end{figure}

Verification of the period value is certainly required. It may be difficult to gather a significantly better X-ray dataset in the future (Tr16-22 having already been observed 28 times!), but the analysis of spectropolarimetric, photometric, or spectroscopic data could help identifying the rotational timescale. The new spectropolarimetric data yield longitudinal fields of $-447\pm74$\,G ($-77\pm69$\,G for the null diagnostics) for April 2013 and $-517\pm116$\,G ($35\pm114$\,G for the null diagnostics) for July 2013. These values are comparable to the longitudinal fields of $-454\pm72$\,G and $-604\pm87$\,G measured on two consecutive nights of April 2011 \citep{naz12carina}, i.e. the field strength appears stable. Furthermore, the H$\beta$ and H$\gamma$ line profiles are similar in all FORS datasets (Fig. \ref{balmer}) and they also agree well with those in an archival FEROS spectrum from March 2011. Are these results compatible with the potential period of 54d suggested by the X-rays? In a magnetic oblique rotator, as the star rotates, the optical emissions, X-ray luminosity, and recorded longitudinal field are expected to vary. This simultaneity is indeed observed for the two magnetic O-stars with X-ray monitorings, HD\,191612 and $\theta^1$\,Ori\,C \citep{gag05,naz10}. In fact, the observed constancy of the optical data of Tr16-22 is fully compatible with the 54d X-ray period because all the optical data would have actually been obtained at similar phases. Folding observing dates with the X-ray derived period and choosing $T_0$ as the date of the oldest \xmm\ observation (\#4 in Table \ref{list}, $JD$=2\,451\,751.707), the phases are 0.52 for the FEROS observation, 0.34 for the 2011 FORS data, 0.45 and 0.34 for the new FORS data of April and July 2013, respectively. This corresponds to the maximum emission phase in X-rays, and may well explain why no large variations are seen in the spectroscopic and spectropolarimetric observations. The alternative periods mentioned above provide greater spread in the phases of the longitudinal magnetic field measurements  ($\Delta(\phi)\sim 0.4$ or 0.6 rather than 0.2), making it harder to explain the comparable strength of the field at the two epochs of observation.

To try to confirm the period, we have also examined the photometry of Tr16-22. Available photometric data of Tr16-22 were taken in the framework of the All Sky Automated Survey\footnote{http://www.astrouw.edu.pl/asas/?page=main}, where the star is designed by \#104508-5946.1, 104508-5946.2 and 104511-5946.1). Keeping only the data with best quality (flag `A'), we found that (1) some data taken before JD=2\,453\,500 show a great scatter (up to 0.75 mag deviation from the mean), hence are probably erroneous, and (2) no significant periodicity can be identified in the reliable data taken after that date. This is certainly due to the large error bars ($\sim$0.05\,mag) compared to the low-amplitude photometric variations of magnetic OB stars (a few tens of mmag in favourable cases, e.g. \citealt{koe02,bar07}).

\section{Conclusion}

Using a large dataset of \xmm\ and \ch\ observations, we have analyzed the high-energy behavior of the magnetic massive star Tr16-22. This star displays simultaneous flux and hardness variations. They appear to be recurrent with a timescale of $\sim$54d, making Tr16-22 the third magnetic O-star known to present X-ray variations. In the magnetic oblique rotator framework, this timescale represents the rotational period of the star. The rotation appears quite slow, but that agrees both with observations of other magnetic O-stars \citep[see][and references therein]{pet13} and expectations from magnetic braking \citep{udd09}. The simultaneous hardness changes imply that not all X-ray components are occulted in the same way, suggesting some stratification within the confined winds zone. 

Two additional spectropolarimetric datasets confirm the magnetic nature of the star. The derived values of the longitudinal fields are similar to those observed two years before, but it should be noted that all optical datasets were likely taken at similar phases considering the X-ray period. Further spectral and spectropolarimetric monitorings are now needed, to derive the magnetic geometry of Tr16-22 and enable its modeling.

\begin{acknowledgements}
YN acknowledges support from  the Fonds National de la Recherche Scientifique (Belgium), the Communaut\'e Fran\c caise de Belgique, the PRODEX XMM and Integral contracts, and the `Action de Recherche Concert\'ee' (CFWB-Acad\'emie Wallonie Europe). VP acknowledges support from NASA through Chandra Award numbers G02-13014X and TM-15001C issued by the Chandra X-ray Observatory Center which is operated by the Smithsonian Astrophysical Observatory for and behalf of NASA under contract NAS8-03060. GAW acknowledge Discovery Grant support from the Natural Science and Engineering Research Council of Canada (NSERC). ADS and CDS were used for preparing this document. 
\end{acknowledgements}

\end{document}